\documentclass[preprint2]{aastex}
\shorttitle{Near-Infrared studies of RS Ophiuchi}
\shortauthors{Das, Banerjee $\&$ Ashok}
\begin{document}
\title{A near-infrared shock wave in the 2006 outburst of recurrent nova RS Ophiuchi}

\author{Ramkrishna Das, Dipankar P.K. Banerjee and Nagarhalli M. Ashok}
\affil{Physical Research Laboratory, Navrangpura,  Ahmedabad,  
Gujarat 380009, India}
\email{rkdas,orion,ashok@prl.res.in}

\begin{abstract}
Near-infrared spectra are presented for the recent 2006 outburst of the
recurrent nova RS Ophiuchi (RS Oph).We report the rare detection of an infrared 
shock wave  as the nova ejecta  plows into the pre-existing wind 
of the secondary in the RS Oph system consisting of a white dwarf (WD) primary 
and a red giant secondary. The evolution of the shock is traced through 
a free expansion stage to a decelerative phase. The behavior of
the shock velocity with time is found to be broadly consistent with
current shock models. The present observations also imply that the WD in the RS Oph system
has a high mass indicating that it could be a potential SNIa candidate. We also
discuss the results from a recent study showing that the near-IR 
continuum from the recent RS Oph eruption does not originate in an expanding fireball. However, the present work shows that the IR line emission  does 
have an origin in an expanding shock wave.

\end{abstract}


\section{Introduction}
RS Ophiuchi, the well known recurrent nova (RN), 
underwent its sixth recorded outburst recently on 2006 February 12. The
five previous known eruptions of the object occurred in 1898, 1933, 1958, 1967 and
1985. The binary components of the RS Oph system consist of a massive
white dwarf and a M8III giant  with the orbital period of
the system being 460 days (Fekel et al. 2000). While the outburst of a 
recurrent and a
classical nova share a common origin in a thermonuclear runaway on a WD
surface which has accreted matter from a companion star, an important
post-outburst distinguishing feature between the two is that the high velocity
ejecta  in a recurrent nova outburst is immediately impeded by the
surrounding wind of the red giant companion. This leads to the
generation of a shock wave that propagates into the red giant wind. The
temporal evolution of the shock and associated physical parameters like the
shock velocity have been predicted theoretically but an astrophysical
environment where such predictions can be rigorously tested are
extremely rare to come by. Recurrent novae provide such an opportunity
for testing and therefore have special significance. Another class of
objects where similar shock waves are open to study are young supernova
remnants  but here the evolution of the shock occurs on the timescale of
a few hundreds of years (vis-a-vis a few days for an analogous
development to occur in a RN) making it difficult to follow the
evolution in supernova
remnants.

The recent outburst of RS Ophiuchi has been extensively studied at
different wavelength regimes viz. in X-rays (Sokoloski et al. 2006;
Bode et al. 2006), in the radio (O'Brien et al. 2006) in the infrared
(Evans et al. 2006; Monnier et al. 2006) and in the optical (Iijima 2006; 
Buil 2006; Fujii 2006). The X-ray results clearly
detect a X-ray blast wave that expands into the red giant wind. The
XRT observations from the Swift satellite (Bode et al. 2006) trace the
temporal change in the shock velocity based on a set of 8 observations
between 3.16 to 26 days after the outburst. Similarly the Rossi X-ray
Timing Explorer (RXTE) observations (Sokoloski et al. 2006) cover the
shock wave evolution based on 6 epochs of observations between 3 days
to 21 days after the eruption. In context of the above results, the
current observations present some new aspects in addition to having the
advantage of a  more comprehensive temporal coverage.  The major
new result lies in our detection of an infrared shock wave which
manifests itself through the narrowing of emission lines observed in
the spectra. The detection of such a shock wave in the infrared has not been
made earlier in a RN. The present data, which cover 23 epochs of
observation between 1 to 47 days after outburst, also have
significantly better sampling, especially in  the $\sim$ first 20 days
after outburst when fairly fast changes are seen in the shock evolution.
Additionally, the data cover a much more extended period. Thus this
appears to be the most comprehensive data set available for analysing
the shock evolution in RS Oph-like systems and for validating related 
theoretical models. While we make such an
analysis, based on the theoretical shock model of Bode $\&$ Kahn (1985), some
deviations that are seen between model and observations suggest the
data set could be invaluable for testing alternative models that may be
proposed subsequently or for introducing refinements/modifications in
presently available ones. We also use our observations to discuss	 the 
recent result of Monnier et al. (2006) which rules
out the origin of the near-IR emission to be from hot gas in an
expanding shock.

\section{Observations}
 Since its outburst in February 2006, near-IR $JHK$ spectra of RS Oph have been
 taken extensively from the 1.2m telescope at the Mt. Abu Observatory. An
 early set of spectra, taken soon after the outburst, is described in Das, Ashok
 $\&$ Banerjee (2006). In the present work, we concentrate on the $J$ band spectra
 with the primary aim of establishing the detection of a shock wave in the
 near-infrared. A more detailed analysis of other aspects of the $J$ band data and
 also the remaining $H$ and $K$ band data - involving both spectroscopy and photometry
 results -   will be 
 addressed in a future work.  The $J$ band spectra presented here 
 were obtained at a dispersion of 9.5 {\AA}/pixel using a Near Infrared 
 Imager/Spectrometer with a 256$\times$256 HgCdTe NICMOS3 array. 
 Generally, a set of at least two spectra were taken 
 with the  object dithered to  two  different positions along the  slit.   
 The spectrum of the comparison star  SAO 122754  (spectral type A0V), after removing the 
 hydrogen absorption lines in its spectrum by interpolation,  was used
 to ratio the spectra of RS Oph. The object and the comparison star were 
 always observed at similar air-mass to ensure the ratioing process reliably removes 
the telluric lines  in the RS Oph spectra. Wavelength calibration   was done 
using OH sky lines  that register with the spectra and spectral reduction and 
analysis  were done  using IRAF tasks. The observational details are presented 
in Table 1 - the outburst date is assumed to be 12.83 Feb 2006 
(Hirosawa 2006)

\section{Results}
      
The J band spectra are shown in Figure 1. The prominent lines seen are 
HeI (1.0833${\rm{\mu}}$m) and  Paschen $\gamma$ (1.0938${\rm{\mu}}$m) - these two lines are blended in the early spectra, OI lines at 1.1287${\rm{\mu}}$m and 1.3164${\rm{\mu}}$m and Paschen $\beta$ at 1.2818${\rm{\mu}}$m. The presence of a few weaker lines, seen later after the outburst, will be discussed in a future paper.  All the
emission lines strikingly show a narrowing with time - a phenomenon also
seen during in the UV during the 1985 outburst (Shore et al. 1996). Such 
a behavior  implies a reduction
in expansion velocity best explained by associating the
line emitting matter with  a decelerating shock wave. The propagation of the shock wave
in RS Oph-like systems, assuming a spherical geometry, has been studied earlier and can be divided into three 
phases (Bode $\&$ Kahn  1985). In Phase 1 - or the ejecta dominated stage - the ejecta expands freely into the red giant wind and produces a shock at constant velocity. 
In Phase 2, the shock wave is driven into the wind and the shocked material  is
so hot that there is negligible cooling by radiation losses - hence called the 
adiabatic phase. During this phase a deceleration is seen in the shock whose velocity v${_{\rm s}}$ versus time t is expected to behave as
v${_{\rm s}}$ $\alpha$  t${^{\rm -1/3}}$ assuming a r${^{\rm -2}}$ dependence for the decrease in density of the 
wind. In Phase 3, the shocked material has cooled by radiation and here
the expected dependence of the shock velocity is v${_{\rm s}}$ $\alpha$ t${^{\rm -1/2}}$. 
To confirm whether these dependencies are indeed observed, we select the Pa$\beta$ 1.2818${\rm{\mu}}$m and the  OI 1.1287${\rm{\mu}}$m lines
for detailed analysis.   These lines  are strong yet unblended 
by other lines (unlike the HeI 1.0833${\rm{\mu}}$m and Pa$\gamma$ 1.0938${\rm{\mu}}$m lines) ensuring a reliable estimate of their widths. We interpret the observed line widths to be due to kinematic broadening i.e. caused by the dispersion  of the line-of-sight 
velocity component from matter at different parts of an expanding shell (which we assume to be spherical). In this 
case,  the FWHM and FWZI (full width at half maximum and zero intensity respectively) are good indicators of the expansion velocity (i.e. the 
shock velocity). In particular, half the FWZI should be a reasonably good measure of 
the expansion velocity since the largest blue-shifted and red-shifted velocities seen 
in the FWZI come from those parts of the shell that are directly approaching or 
receding from the observer. We present the observed line widths in Table 2. Since 
the observed FWHM of the lines at later stages (beyond March 20) is comparable to the 
FWHM of the instrument profile (which is well approximated by a Gaussian with FWHM = 450 
km/s; FWZI = 1350 km/s) we have deconvolved for the instrument broadening. For this, we 
have used the simple, but sufficiently adequate, approximation that the square of the 
observed FWHM equals the sum of the squares of the instrument and true FWHM's (and 
similarly so for the FWZI).  This assumes a gaussian nature for the observed profiles
which we find to be reasonably valid. The effect of the deconvolution is found to be rather small for the observations prior to 20 March which compose the bulk of the data. 
We have also considered whether optical depth effects could have
affected the line profiles and hence their widths. This appears unlikely especially for
the hydrogen recombination lines. The results of Evans et al. (2006) and also our
H band data (to be presented elsewhere) show that the strength of the hydrogen lines  
can be approximated by Case B predicted strengths indicating that they are optically 
thin. About the OI line we cannot be certain, but its column density is expected
to be several orders lower than that of hydrogen (by virtue of a lower abundance) and 
hence it is unlikely to be optically thick when the hydrogen lines are not so. We show 
the temporal variation of the deconvolved widths of the selected lines with time in 
Figure 2. \\

Figure 2 first establishes that there is indeed a stage of free expansion - lasting 
approximately 4 days - during which the shock velocity remains approximately constant.
This behavior is more evident in the OI data vis-a-vis the Pa$\beta$ data which 
suggests there
may be a mild deceleration during this phase. Following Phase 1, a deceleration
is seen and since this slowing down occurs only after the ejecta has swept up mass from 
the red 
giant wind  comparable to the ejecta mass, the onset of deceleration can be used 
to estimate the mass of the ejected shell. Following Sokoloski et al. (2006), we 
assume the density in the binary to
be 10${^{\rm 9}}$ cm${^{\rm -3}}$ from constraints on the mass-loss rate of the red 
giant (Dobrzycka $\&$ Kenyon 1994) and adopt a mean initial
expansion velocity in the range 3500-4000 km/s based on our results and optical
reports (Buil 2006; Fujii 2006). Thus by day 4, the swept up mass (or equivalently the
ejected shell mass) is estimated to have a mean value of 
$\sim$ 3x10${^{\rm -6}}$$M$$_\odot$. 
This value  reasonably matches with  the Hachisu $\&$ Kato (2001) model for RS Oph 
which estimates that an ejecta mass
of $\sim$ 2x10${^{\rm -6}}$$M$$_\odot$ needs  a mass of 1.35$M$$_\odot$ for the
 white dwarf. We  therefore independently infer that  RS Oph certainly contains a 
 massive WD with a mass close to 1.35$M$$_\odot$ 
- a similar conclusion is reached from alternative methods viz. from a lightcurve 
analysis (Hachisu $\&$ Kato 2001) and from X-ray observations (Solokoski et al. 2006).\\

We find that the decelerative phase, for both the OI and Pa$\beta$ lines is reasonably
well reproduced by power law fits of the form  t${^{\rm -\alpha}}$ with $\alpha$
varying between 0.45 to 0.79.  Using a nonlinear regression 
technique, the best fits were determined by maximizing the correlation coefficient $R$  with a value of $R$ = 1 describing a perfect fit. The plots in Figure 2 have $R$ values of 0.977, 0.988, 0.976 and 0.975 for the FWZI and FWZM of the
Pa$\beta$ and OI lines respectively.   A mean  value of $\alpha$ = 0.64 (with
a standard deviation of 0.14 ) can be said to characterize the  data which is consistent with the Swift results (Bode 
et al. 2006) wherein a value of $\alpha$ = 0.6 is obtained. We do note  
that the results in Figure 2 do suggest the need of using more than 
one power law for different segments. However, our present aim is to obtain a simple 
yet reasonable assessment
of   the overall behavior of the line-widths and therefore we elect to use a single 
power law. We note  that there seems to be a lack - or a very short-lived duration of 
 Phase II (the 
Sedov-Taylor phase) in our data. While the reason for this is not clear, it is possible
that the cooling of the ejecta begins very quickly. We find that the mean value of 
$\alpha$ = 0.64 compares better to 
the  expected value  of 0.5 for Phase III rather than 0.33 expected in Phase II 
(Bode $\&$ Kahn 1985). Thus there are deviations in the observed data from 
expected model predictions. It is possible that a part of such variations arise from 
the assumption of spherical symmetry that is being invoked. But as the radio imaging
(O'Brien et al. 2006) and interferometric results (Monnier et al. 2006) show, the
ejecta remnant has deviations from spherical symmetry and also jet-like structures. 
Such variations in geometry are expected to affect the  observed line profiles.
A treatment of these effects is expected to be involved and beyond the scope of the 
present work. However, we find that on the whole, our results are broadly consistent 
with the shock model wherein
a free expansion stage is expected to be seen followed by a decelerative stage 
characterized by a power law decay in the shock velocity. We thus believe that 
the observations clearly support the presence of an infrared shock wave  
in the recent outburst of RS Oph.\\

While the near-IR line emission, as seen from the 
Pa$\beta$ and OI line behavior, is associated with a moving shock, the site of its
emission within the shock structure needs to be examined. The structure
of the shock is understood to consist of a cool contact surface (i.e. the region of 
discontinuity) positioned
between the hot compressed region consisting
of the ejecta material and the hot swept-up red giant wind (e.g. Lamers $\&$ Cassinelli 
1999). The last two regions are
the sites where the shocked gas is heated to the extremely high  temperature of 
10${^{\rm 8}}$K (as determined for RS Oph; Sokoloski et al. 2006; Bode at al. 2006) and 
which is the source of the observed
X-ray emission. Clearly the IR line emission
cannot emanate from these zones because the presence of neutral OI atoms
in such a hot region is not possible. Thus we think that the near-IR emission
originates from the contact surface, which is considerably cooler and denser, but which 
still propagates with the same velocity as the shock front (Lamers $\&$ Cassinelli 1999)
thereby leading to similar 
kinematic behavior for the shock velocity as predicted for the X-rays 
(Bode $\&$ Kahn 1985).

\section{Discussion}  
Monnier et al. (2006) measured the near-infrared (H and K band) size of RS Oph during 
the recent
outburst using interferometry to show that the object displays
a near-constant size of $\sim$ 3 milli-arcseconds (mas) over the first 60 days of the 
outburst.
At a distance of 1.6 kpc to RS Oph, the near-IR emission 
is expected to expand at the rate of 1 mas per day if associated with an 
expanding shock. Since this is not seen,  these authors are led to the interesting 
conclusion  that 
the IR continuum  does not arise from an expanding fireball - a scenario favored 
to explain  novae outbursts. In context of this, we note that the  net 
IR emission consists of two parts - the continuum and the line emission. Regarding the 
line emission, the present
results clearly show that it  originates from
an expanding shock. This does not resolve the extent and site of origin of the near-IR 
continuum i.e. what part could  arise from the expanding shock wave vis-a-vis the 
fraction that arises from other sites/mechanisms (e.g. the  alternative theory of 
Hachisu 
$\&$ Kato 2001). But an aspect of the Monnier et al. (2006) data, which maybe worthwhile
to reexamine, is whether they do detect a faint signature associated with
the expanding shock front. The reason to expect such a detection  is that our H band 
spectra (also see the Evans et al. 2006 spectra), 
taken contemporaneously with Monnier et al. (2006), are replete with strong Brackett
series lines. The combined flux contribution  from these lines is estimated to be 
$\sim$ 15-20 
percent relative to the continuum. It would therefore  appear 
likely that some signature  of this line emission, though weaker than the continuum,
 should be detected and  show up in expansion in the interferometric data.

\begin{acknowledgements}
Research at the Physical Research Laboratory  is funded by the Department of 
Space, Government of India.  	  
	  
\end{acknowledgements}

%______________________________________________________________

\clearpage

\begin{figure}
\epsscale{2.0}
\plotone{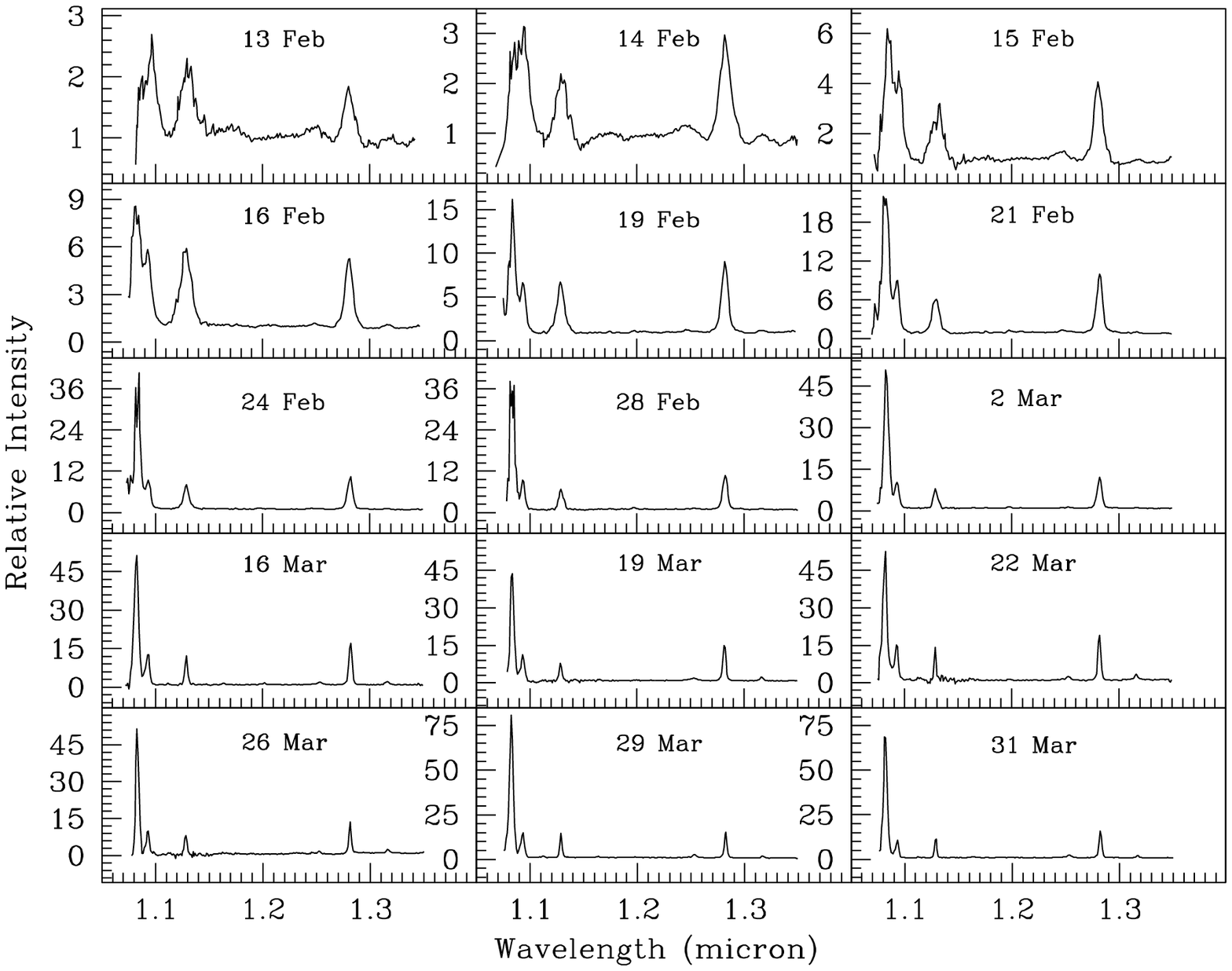}
\caption{Selected near-IR $J$ band spectra of the 2006 outburst of RS Oph are shown between 1 
to 47 days after outburst. A few of the spectra listed in Table 1 are not included in 
the figure - such spectra being rather similar to those preceding or succeeding them in 
sequence of time. The prominent lines seen are 
HeI (1.0833${\rm{\mu}}$m) and  Paschen $\gamma$ (1.0938${\rm{\mu}}$m) - these two lines are blended in the early spectra but resolved at later stages, the OI line at 1.1287${\rm{\mu}}$m 
and Paschen $\beta$ at 1.2818${\rm{\mu}}$m. \label{fig3}}
\end{figure}

\clearpage

\begin{figure}
\plotone{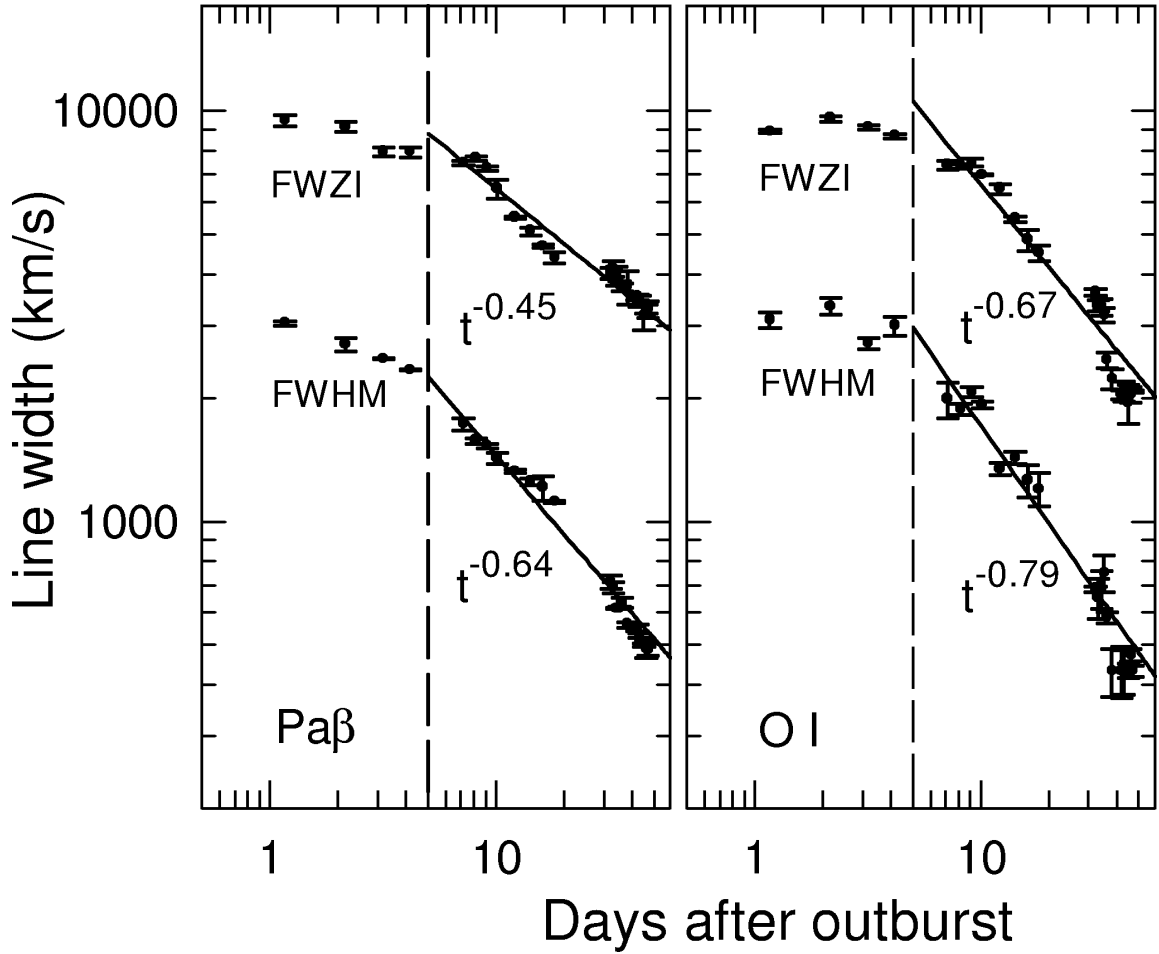}
\caption{The behavior of the deconvolved line widths for the Pa$\beta$
 1.2818${\rm{\mu}}$m and the OI 1.1287${\rm{\mu}}$m lines is shown.
The line widths imply a free expansion phase for the infrared shock front for the first
four days - the region to the left of the dashed drop lines. This phase is followed by a decelerative phase which is 
best described by a power law decline in the shock velocity with time. 
The different power laws used for fitting are marked in the figure; data points are shown with 1$\sigma$ error bars. Further details can 
be found in the text in Section 3. \label{fig1}}
\end{figure}

\clearpage

\begin{table}
\caption{Log of  J band spectral observations for RS Oph }
\begin{tabular}{cccccc}
\hline 
\hline\\
Date (UT)         & Days after &Int.    &  Date(UT)           & Days after   &Int.    \\    
2006              & outburst   &time(s) &  2006               & outburst     &time (s)\\      
\hline      
Feb 13.9929       &  1.1629    &10      &   Mar 16.9465       & 32.1165      & 20     \\
Feb 14.9915       &  2.1615    &10      &   Mar 17.8866       & 33.0566      & 30     \\
Feb 15.9898       &  3.1598    &10      &   Mar 18.9372       & 34.1072      & 30     \\
Feb 16.9866       &  4.1566    &10      &   Mar 19.8833       & 35.0533      & 30     \\ 
Feb 19.9759       &  7.1459    &10      &   Mar 20.8733       & 36.0433      & 45     \\
Feb 20.9721       &  8.1421    & 5      &   Mar 22.8682       & 38.0382      & 45     \\
Feb 21.9463       &  9.1163    &10      &   Mar 26.8778       & 42.0478      & 60     \\
Feb 22.9505       & 10.1205    &10      &   Mar 27.8644       & 43.0344      & 60     \\
Feb 24.9705       & 12.1405    &10      &   Mar 29.8462       & 45.0162      & 75     \\
Feb 26.9926       & 14.1626    &10      &   Mar 30.8629       & 46.0329      & 75     \\
Feb 28.9542       & 16.1242    &15      &   Mar 31.8701       & 47.0401      & 90     \\
Mar  2.9465       & 18.1165    &20      &                     &              &        \\

\hline
\hline\\
\end{tabular} 
\end{table}

\clearpage

\begin{table}
\caption{Observed widths of the Pa$\beta$ and OI lines}
\begin{tabular}{ccccccccccc}
\hline 
\hline\\
Days       &  Pa$\beta$   & Pa$\beta$ & OI & OI          &Days     & Pa$\beta$  & Pa$\beta$ & OI &OI &   \\ 
after      &  fwhm & fwzi      &fwhm  & fwzi          &after    & fwhm  & fwzi & fwhm & fwzi     &   \\
outburst   & km/s  & km/s & km/s   &km/s  &outburst &km/s     &km/s& km/s  &km/s&\\ 
\hline      
 1.16      & 3066  & 9561      & 3136 & 9009      & 32.12   & 839   & 4236    & 850  & 3853 &    \\
 2.16      & 2738  & 9228      & 3375 & 9647      & 33.06   & 822   & 4330    & 824  & 3614 &    \\
 3.16      & 2532  & 8051      & 2764 & 9221      & 34.11   & 758   & 4189    & 850  & 3588 &    \\
 4.16      & 2381  & 8063      & 3029 & 8770      & 35.05   & 759   & 4026    & 904  & 3455 &    \\
 7.15      & 1782  & 7595      & 2046 & 7494      & 36.04   & 775   & 3955    & 771  & 2817 &    \\
 8.14      & 1632  & 7759      & 1940 & 7441      & 38.04   & 715   & 3961    & 664  & 2604 &    \\
 9.12      & 1587  & 7361      & 2126 & 7600      & 42.05   & 708   & 3768    & 664  & 2445 &    \\
10.12      & 1489  & 6595      & 1993 & 7122      & 43.03   & 690   & 3698    & 664  & 2471 &    \\
12.14      & 1398  & 5648      & 1435 & 6590      & 45.02   & 669   & 3452    & 664  & 2365 &    \\
14.16      & 1326  & 5266      & 1515 & 5607      & 46.03   & 660   & 3557    & 691  & 2445 &    \\
16.12      & 1287  & 4868      & 1355 & 5023      & 47.04   & 659   & 3557    & 664  & 2471 &    \\
18.12      & 1203  & 4587      & 1302 & 4704      &         &       &         &      &      &    \\
%%&&&&\\
\hline
\hline\\
\end{tabular} 
\end{table}

\end{document}